\documentclass[fleqn,twoside]{article}
\usepackage{espcrc2}
\usepackage{bm}
\usepackage{graphicx}

\newcommand{\bfu}{\bm{u}}
\newcommand{\Nc}{N_{\rm cor}}
\newcommand{\Ne}{N_{\rm exp}}
\title{Bayesian curve fitting for lattice gauge theorists
}

\author{
C.\ Morningstar\address[CMU]{Physics Department, 
    Carnegie Mellon University, Pittsburgh, PA, USA}
}
       
\begin{document}

\begin{abstract}
A new method of extracting the low-lying energy spectrum
from Monte Carlo estimates of Euclidean-space correlation functions
which incorporates Bayesian inference is described and tested.
The procedure fully exploits the information present in the correlation
functions at small temporal separations and uses this information in a
way consistent with fundamental probabilistic hypotheses.  The computed
errors on the best-fit energies include both statistical uncertainties
and systematic errors associated with the treatment of contamination
from higher-lying stationary states. Difficulties in performing the
integrals needed to compute these error estimates are briefly discussed.
\end{abstract}

\maketitle

\section{INTRODUCTION}

The physical consequences of a quantum field theory can be deduced
from the correlation functions of the fields.  Usually this is done by
matching the correlation functions (or suitable combinations of them) to
known behavior which contains the observables of interest.
The path integrals which yield the correlation functions are often
computed using Monte Carlo estimates.

To estimate the low-lying energy spectrum of the stationary states of
a theory, one studies a {\em matrix} of correlation functions 
\begin{equation}
  h^{ij}(t) = \sum_{t_0}\langle\ \phi_i(t\!+\!t_0)
 \ \phi^\dagger_j(t_0)\ \rangle,
\end{equation}
where $\langle \ \phi_i(t)\ \rangle=0$ and $\phi_i^\dagger(t)$ are
operators capable of creating the states of interest. As the temporal
separation $t$ becomes large, the elements of this correlation matrix
tend (neglecting boundary conditions) to a handful of damped exponentials
whose decay rates yield the lowest-lying energies of the system.
To reliably determine these decay rates, the correlators $h^{ij}(t)$ must
be calculated for large enough $t$ such that they are well approximated by
only small numbers of exponentials.  The Monte Carlo 
estimates of the correlation matrix $h^{ij}_D(t)$, where now $D$
denotes Monte Carlo estimates, must then be fit 
to known asymptotic forms $h^{ij}_M(t)$,
where $M$ denotes the model fitting functions.  The fit is carried out
by adjusting the fit parameters, collectively denoted by $\bfu$, in
$h^{ij}_M(t)$ in order to maximize
the likelihood of finding the results $h^{ij}_D(t)$ given $h^{ij}_M(t)$.
This amounts to minimizing
\begin{equation}
\chi^2 = \sum_{\alpha\beta} \Bigl( M_\alpha(\bfu)\!-\!D_\alpha\Bigr)
 \ C^{-1}_{\alpha\beta}\  \Bigl( M_\beta(\bfu)\!-\!D_\beta\Bigr),
\label{eq:chisq}
\end{equation}
where $M_\alpha(\bfu)$ refers to $h^{ij}_M(t)$ and $D_\alpha$ refers to
$h^{ij}_D(t)$.  The indices $\alpha$ and $\beta$ each include the
indices of the correlation matrix as well as the instants in time at
which the correlation matrix is sampled.  $C_{\alpha\beta}$ is the
covariance matrix associated with $D_\alpha$.  One commonly-used method is
to use the fitting function (which neglects boundary conditions)
\begin{equation}
h_M^{ij}(t) = \sum_{p=0}^{n-1} \ Z^{ip}Z^{jp} \exp(-tE_p),
\label{eq:asym1}
\end{equation}
with $E_{p+1}>E_p$, 
and fit the $n\!\times\!n$ matrix $h_D^{ij}(t)$ for $t_{\rm min}\leq t
\leq t_{\rm max}$.  If the $n$ operators have been constructed using
a variational approach applied to a much larger set of operators,
the off-diagonal matrix elements $h^{i\neq j}_D(t)$
are often statistically consistent with zero.  In such cases, little
is gained by including these correlation functions in the fitting,
and hence, one may fit only to the $n$ diagonal correlators using
the fitting function
\begin{equation}
h_M^{(i)}(t) = \sum_{p=0}^{n-1} \ Z^{ip} \exp(-tE_p),
\label{eq:asym2}
\end{equation}
where now the $Z^{ip}$ are usually real and positive. 

\begin{figure}[t]
\includegraphics[width=3.0in,bb=18 204 592 718]{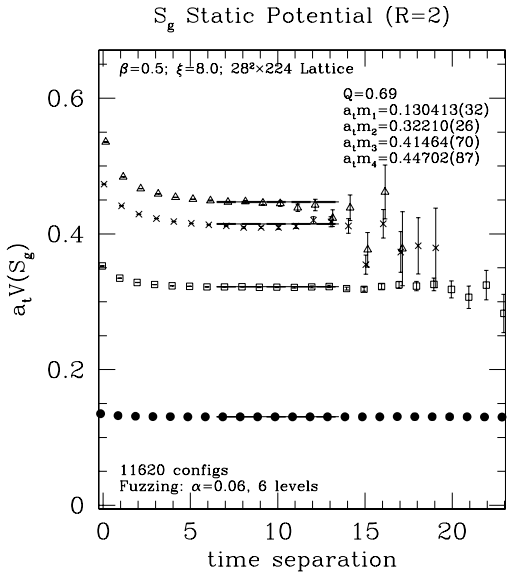}
\caption[cap]{Effective mass plot showing fits for the lowest four energies
of Example 1 using the model fitting function from 
Eq.~(\protect\ref{eq:asym2}) and $7\leq t\leq 14$.  The data points in this
figure are effective masses as defined by Eq.~(5.18) in 
Ref.~\cite{luschereffmass} such that the lowest $n$ effective masses tend
to the lowest $n$ physical masses as the temporal separation becomes
large.  Note that the correlation functions themselves are used in
the fitting, not the effective mass points which are included only
for illustrative purposes.
\label{fig:example1A}}
\end{figure}

\begin{table}[t]
\caption[cap]{Estimates for the lowest-lying energies $E_0$ and $E_1$
from the standard analysis applied to Example 1.  The fits are determined
for times $t$ satisfying $t_{\rm min}\leq t\leq t_{\rm max}$ and
$Q$ indicates the fit quality.  The errors shown on $E_0$ and $E_1$ are
determined using the bootstrap method.
\label{tab:one}}
\begin{center}
\renewcommand{\tabcolsep}{1.5mm} 
\begin{tabular}{ccccc}\hline
$t_{\rm min}$ & $t_{\rm max}$ & $Q$ & $E_0$ & $E_1$\\ \hline
   3 &  10 &     0     &  0.130632(27) & 0.32435(10)\\
   4 &  11 &     0     &  0.130514(24) & 0.32319(10)\\
   5 &  12 &     0     &  0.130449(27) & 0.32268(14)\\
   6 &  13 &     0.23  &  0.130421(33) & 0.32217(21)\\
   7 &  14 &     0.69  &  0.130413(32) & 0.32210(26)\\
   8 &  15 &     0.21  &  0.130404(31) & 0.32206(32)\\ \hline
\end{tabular}
\end{center}
\end{table}

When using either Eq.~(\ref{eq:asym1}) or (\ref{eq:asym2}) as the fitting
function, it is crucial that $t_{\rm min}$ be chosen such that
contributions from all higher energy eigenstates are negligible.
Generally one is guided by the quality of fit to determine this,
but the guidelines for doing so are not very clear.
For example, one could start with $t_{\rm min}=0$, for which the
fit quality is usually very bad, and increase $t_{\rm min}$ by one
temporal lattice spacing at a time, keeping $t_{\rm max}=t_{\rm min}
+t_{\rm window}$ for fixed $t_{\rm window}$, until some acceptable fit
quality, such as $Q=0.2$, is achieved.  But the acceptable value of $Q$
and the value of $t_{\rm window}$ are somewhat subjective, and it is
not clear how to incorporate the uncertainties in choosing these values
into the final best-fit values for the energy levels.  More stringent
values of $Q_{\rm acc}$ and $t_{\rm window}$ generally lead to larger
values of $t_{\rm min}$, and hence, larger errors in the best-fit
energies.  The subjectivity of these errors and the loss of information
associated with the discarded $t<t_{\rm min}$ measurements of the
correlators are undesirable features of this approach.

\begin{figure}[t]
\includegraphics[width=3.0in,bb=18 204 592 718]{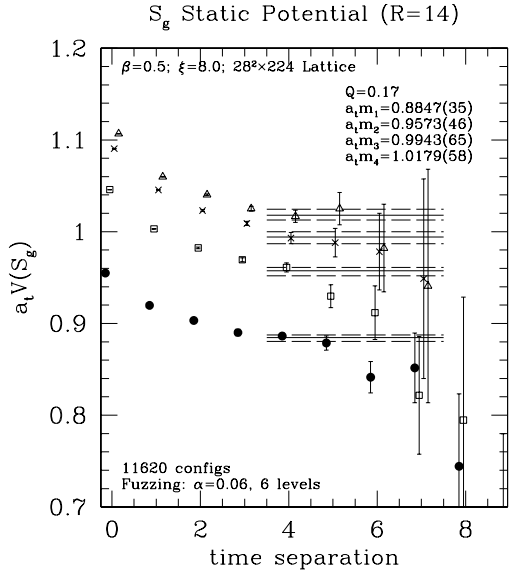}
\caption[cap]{Effective mass plot showing fits for the lowest four energies
of Example 2 using the model fitting function from 
Eq.~(\protect\ref{eq:asym2}) and $4\leq t\leq 8$. Other details in
this figure are the same as in Fig.~\protect\ref{fig:example1A}.
\label{fig:example2A}}
\end{figure}

\begin{figure}[t]
\includegraphics[width=3.0in,bb=18 204 592 718]{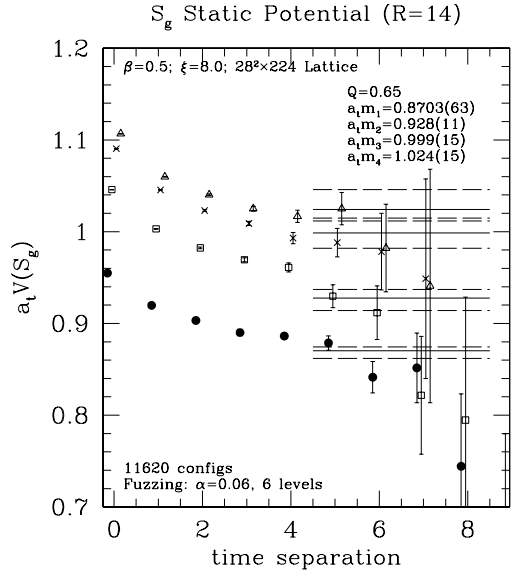}
\caption[cap]{Effective masses with fits for $5\leq t\leq 9$. See
Fig.~\protect\ref{fig:example2A} for other details.
\label{fig:example2B}}
\end{figure}

For illustrative purposes, consider two examples, one with
an excellent signal (Example 1), the other with a questionable signal
(Example 2).  Both examples are energies of a gauge field in the
presence of a static positive and a static negative charge in
2+1-dimensional compact U(1) lattice gauge theory using an improved
action on an anisotropic lattice.  Typical fit results for the
lowest few energies using the standard analysis with the fitting
function from Eq.~(\ref{eq:asym2}) are shown in 
Figures~\ref{fig:example1A}-\ref{fig:example2B}.  
$S_g$ denotes the chosen symmetry channel and $R$ is the number of lattice
sites between the charges.  Tables~\ref{tab:one} and \ref{tab:two} detail
the fit results for the lowest-lying energies $E_0$ and $E_1$ in each
example.  In Example 1, a suitable choice for $t_{\rm min}$
and $t_{\rm max}$ is easily guided by the fit quality $Q$ and the
final results are reasonably insensitive to minor changes in the fit
range.  In Example 2, one might agonize over 
whether $t_{\rm min}=4$ or $5$ should be used.
The standard analysis provides no means of incorporating such an
uncertainty into the error estimate. 

\begin{table}[b]
\caption[cap]{Estimates for the lowest-lying energies $E_0$ and
$E_1$ from the standard analysis applied to Example 2. 
\label{tab:two}}
\begin{center}
\renewcommand{\tabcolsep}{2mm} 
\begin{tabular}{cccll}\hline
$t_{\rm min}$ & $t_{\rm max}$ & $Q$ & \hspace{5mm}$E_0$
 & \hspace{5mm}$E_1$\\ \hline
   2  &  6  &    0     &  0.90102(78) &  0.98012(91)\\
   3  &  7  &    0.04  &  0.8894(16)  &  0.9685(21)\\
   4  &  8  &    0.17  &  0.8847(35)  &  0.9573(46)\\
   5  &  9  &    0.65  &  0.8703(63)  &  0.928(11)\\
   6  & 10  &    0.52  &  0.832(18)   &  0.895(27)\\
   7  & 11  &    0.64  &  0.799(68)   &  0.828(47)\\ \hline
\end{tabular}
\end{center}
\end{table}

In order to incorporate information from small time separations, many
exponentials must be retained in the model fitting functions.  Standard fits
using many exponentials quickly encounter severe problems: often,
fitting instabilities occur and uncertainties in the estimates of the
parameters of interest become very large.  The source of these
problems is the fact that unconstrained fits allow physically insensible
or impossible values for those parameters which are not adequately
constrained by the data.  The solution to this problem is to introduce
constraints. {\em Bayesian statistics} allows the introduction of such
constraints in a natural way.  The Bayesian approach is now widely used
throughout many disciplines, such as economics, medicine, astrophysics, and
condensed matter physics.

This talk is a report on progress being made in exploring methods of
extracting physical observables from Monte Carlo estimates of
correlation functions which incorporate Bayesian inference into the
statistical analysis.  Such procedures can fully exploit
the information present in the correlation functions at small
temporal separations and use this information in a way consistent
with fundamental probabilistic hypotheses.  The computed errors
on the best-fit parameters include both statistical uncertainties
and systematic errors associated with the treatment of the
excited-state contaminations.  Example 1 will be used to verify
that such new methods work in cases where the signal is excellent.
Example 2 will then be used to see if the new methods provide an
improved means of extracting the information in cases when the
signal is more obscured.  An earlier report on such explorations is
given in Ref.~\cite{peter}.  Bayesian methods have been previously 
discussed in the context of lattice QCD in Ref.~\cite{makovoz},
and more recently, using maximum entropy 
methods in Refs.~\cite{maxentD,maxentA,maxentB,maxentC}.

Since these methods employ Bayesian inference, a brief review of
the Bayesian approach is given in Sec.~\ref{sec:bayes}.  The application
of the Bayesian approach to the particular problem of interest here,
namely, extraction of energies from Monte Carlo determinations of
correlation functions, is then detailed in Sec.~\ref{sec:spectrum}.

\section{BAYESIAN REGRESSION}
\label{sec:bayes}

Bayesian regression is different from the classical (frequentist)
approach.  In the classical approach, the data are the only source
of information explicitly taken into account in constructing an
estimate or test.  In the Bayesian approach, an estimate or test
is produced by combining the current data with information from
past experience and/or theoretical constraints (prior information).
In this section, Bayesian regression is described in general
terms.  Some introductory textbooks on Bayesian statistics
are given in Refs.~\cite{sivia,press,carlin}.  For particular
emphasis on Bayesian regression, see, for example, 
Refs.~\cite{birkes,bretthorst}, and for a few other discussions of
Bayesian methods in physics, see Refs.~\cite{bryan,pdg,agostini}.

The Reverend Thomas Bayes was a Presbyterian minister born in
1702 in London, England, and who died April 17, 1761.  His
theory of probability was published posthumously by Richard Price
in 1763 in an article entitled an {\sl Essay towards solving a
problem in the doctrine  of chances} in the {\sl Philosophical
Transactions of the Royal Society.}  The theorem presented by
Bayes was restricted to the binomial distribution, but its
generality was most likely recognized by Bayes.  Bayes' theorem was
generalized beyond the binomial distribution by Laplace in
1774, most likely without knowledge of Bayes' work.  Standard
(frequentist) statistical methods were developed later than Bayesian
methods. Although linear regression and goodness of fit first appeared
in the late 1800's, the field blossomed during the 1920's with the
works of Fisher, Neyman, and Pearson, with a flurry of research and
applications during World War II.  Bayesian methods are
much older, but were largely ignored or actively opposed until
the 1950's when they were championed by prominent non-statisticians,
such as physicist H.~Jeffreys and economist A.~Bowley.  Their
popularity grew rapidly during the 1970's with the advent of
affordable computers, and heated debates over the superiority
of either method have raged ever since.

The cornerstone of Bayesian statistics is Bayes' theorem:
\begin{equation}
P(M\vert D\cap I) = \frac{ P(D\vert M\cap I)\ P(M\vert I)}{
\int\!dM\ P(D\vert M\cap I)\ P(M\vert I) },
\label{eq:bayes}
\end{equation}
where $D$ represents the data, $M$ denotes the model which one
believes should describe the data, and $I$ represents our prior
knowledge or background information about the system. The conditional
probability distribution $P(D\vert M\cap I)$ is known as the
{\em likelihood} of the data, $P(M\vert I)$ is the {\em prior} probability
distribution, and $P(M\vert D\cap I)$ is called the {\em posterior}
probability distribution.  The term in the denominator of Eq.~(\ref{eq:bayes})
is independent of $M$ and so can often simply be absorbed into an overall
normalization.  Bayes' theorem essentially states that
\[ {\rm posterior}\ \propto\ {\rm likelihood}\ \times \ {\rm prior}.\] 

Bayesian regression uses the {\em posterior} distribution for all statistical
inference.  Model parameters are estimated using one's favorite statistic
but evaluated using the posterior distribution.  Common measures of central
tendency include the {\em mode}, {\em mean}, and {\em median}, and common
measures of dispersion include the {\em variance}, {\em skewness}, and 
{\em kurtosis}.  For example, if the model
is specified by a set of parameters $\bfu$, then the mean value
and variance of a model parameter $u_j$ are given by
\begin{eqnarray}
 \langle u_j\rangle\!\!\! &=&\!\!\! \int\!\! \bm{du}\ u_j\ P(M(\bfu)\vert D\cap I),\\
{\rm var}(u_j)\!\!\! &=&\!\!\! \int\!\! \bm{du}\ (u_j\!-\!\langle u_j\rangle)^2
 \ P(M(\bfu)\vert D\cap I).
\end{eqnarray}
The mode of $u_j$ is found by maximizing the posterior probability, that is,
maximizing the probability of the model $M$ being correct given data $D$ and
subject to the background information $I$.

The likelihood $P(D\vert M\cap I)$ is the probability distribution 
of the finding the data $D$ given the particular set of model parameters
$M$ and background information $I$, and from the Central Limit theorem
 has the form
\begin{equation}
P(D\vert M \cap I) \propto \exp(-\chi^2/2),
\end{equation}
where $\chi^2$ is the familiar form given in Eq.~(\ref{eq:chisq}).

The new and key feature of Bayesian statistics is the {\em prior}
$P(M\vert I)$.  In standard maximum likelihood fits, the Bayesian prior
$P(M\vert I)$ is taken to be a constant and thus, maximizing the likelihood
becomes equivalent to minimizing $\chi^2$.  When the number of fit parameters
is small compared to the number of data points, this is a reasonable thing to
do.  However, when the number of fit parameters becomes comparable or larger
than the number of data points, the above procedure can become unstable
and/or yield parameter estimates with overly large uncertainties, as
mentioned previously.  The role of the Bayesian prior $P(M\vert I)$ is to
filter out parameter values that are impossible and improbable, given
some prior knowledge about the solutions, and hence, introduce constraints
into the fits.

Prior distributions are specified based on information accumulated from
past studies (earlier data), the opinions of subject-area experts, and/or
from theoretical constraints. One often considers logical connections between
the parameters and symmetries which the prior distribution
must obey. To simplify the computational burden, practitioners often restrict
the choice of prior to some familiar distributional form.  In some
cases, the prior can be endowed with very little informative content
(such as maximum entropy which amounts to letting a monkey throw balls
into bins).  Strictly speaking, using the observed data to influence the
choice of prior is against the Bayesian philosophy.  However, it is legitimate
to use a small subset of the data to guide the construction of the prior 
and to use the rest of the data in the regression itself.  It is very
important that one should avoid putting in more information than is truly
known, since results do and should depend on the prior.  The prior is
often viewed both as a nuisance and as an opportunity.

\section{CHOICE OF PRIOR AND TESTING}
\label{sec:spectrum}

By introducing a Bayesian prior, we can now use the model function
of Eq.~(\ref{eq:asym2}) but summing over many more exponentials:
\begin{equation}
h_M^{(i)}(t) = \sum_{p=0}^{\Ne-1} \ A^{(i)}_p \exp(-tE_p),
\label{eq:asym3}
\end{equation}
where $\Ne$ is the number of exponentials, $\Nc$ is
the number of correlators to be simultaneously fit, and 
$i=0,\dots,\Nc\!-\!1$.  To ensure positivity of the coefficients
and to order the energies, define
\begin{equation}
 A^{(i)}_n = \Bigl( b_n^{(i)}\Bigr)^2,\qquad E_n=E_{n-1}+\epsilon_n^2,
\end{equation}
so that the actual parameters used in the fitting are 
$u_\alpha = \{ E_0, \epsilon_n, b_n^{(i)}\}$.  In total, there are
$(\Nc+1)\Ne$ fitting parameters. We then choose a
Gaussian prior
\begin{equation}
P(M\vert I)\propto \exp\left(-\sum_\alpha 
 \frac{(u_\alpha-\eta_\alpha)^2}{2\sigma_\alpha^2}\right).
\label{eq:prior}
\end{equation}
In this work, we shall not include a prior for the first $\Nc$
energies, so that the sum over $\alpha$ in Eq.~(\ref{eq:prior}) includes
all parameters except $E_0,\epsilon_1,\epsilon_2,\dots,\epsilon_{\Nc-1}$.
Hence, there are $\Nc\Ne$ terms in the summation over $\alpha$.
In using such a prior, we incorporate our expectation that the parameter
corresponding to the index $\alpha$ most likely has a value between
$\eta_\alpha-\sigma_\alpha$ and $\eta_\alpha+\sigma_\alpha$.

To simplify matters, we assume that the energies of the excited-state
contamination above the $\Nc$-th level are most likely to be equally spaced.
Also, due to the variational construction of our operators, we expect
that correlator $j$ is dominated by the $E_j$ exponential.  We further
simplify matters by taking the same $\eta_\alpha$ and $\alpha_\alpha$
for all the $b_j^{(j)}$.  All other coefficients are expected to be
small and are given the same expected range of values.
To summarize, our prior is specified using
\begin{equation}
  \eta_\alpha =
 \left\{\begin{array}{lll}
   \varepsilon, &\alpha \Rightarrow\epsilon_j,& j=\Nc\dots\Ne\!-\!1,\\
   \Gamma, &\alpha \Rightarrow b_j^{(j)},& j=0\dots\Nc\!-\!1,\\ 
   \gamma, &\alpha \Rightarrow b_j^{(i)},& i\neq j,
\end{array}\right.\!\!\!\!
\end{equation}
and for the Gaussian widths, we similarly use $\sigma_\varepsilon$,
$\sigma_\Gamma$, and $\sigma_\gamma$.  Choosing values for $\Ne$,
$\varepsilon$, $\Gamma$, $\gamma$, $\sigma_\varepsilon$,
$\sigma_\Gamma$, and $\sigma_\gamma$ completes the specification of
our prior.  $\Ne$ is easily chosen by increasing its value until the
energies of interest stabilize.  The widths must be chosen sufficiently
large so that the fits are not overly constrained beyond our knowledge,
but overly large widths could lead to overly large uncertainties in the
best-fit parameters.  Note that describing the prior is very straightforward,
in contrast to describing how one subjectively chooses fitting ranges
$t_{\rm min}$ and $t_{\rm max}$ in the standard analysis.

\begin{figure}[t]
\includegraphics[width=3.0in,bb=18 204 592 718]{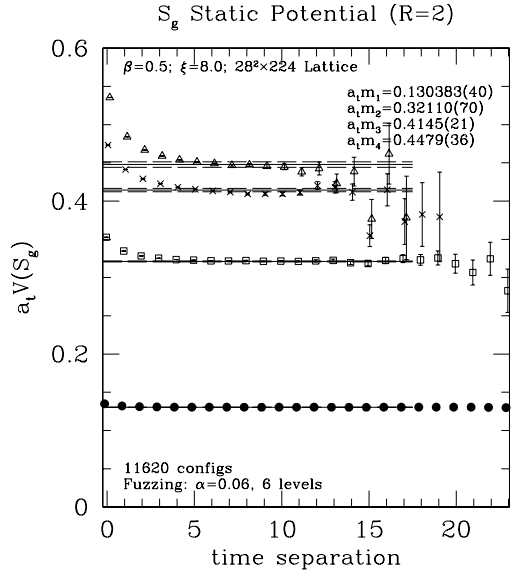}
\caption[cap]{Effective masses and fits for the lowest four energies
of Example 1 using the Bayesian analysis described in 
Sec.~\protect\ref{sec:spectrum} and $0\leq t\leq 18$. 
Prior parameter values are $\Ne=34$, $\varepsilon=0.2$, 
$\sigma_\varepsilon=0.1$, $\Gamma=0.9$, $\sigma_\Gamma=0.2$,
$\gamma=0.05$, $\sigma_\gamma=0.05$.
\label{fig:example1B}}
\end{figure}

\begin{figure}[t]
\includegraphics[width=3.0in,bb=18 204 592 718]{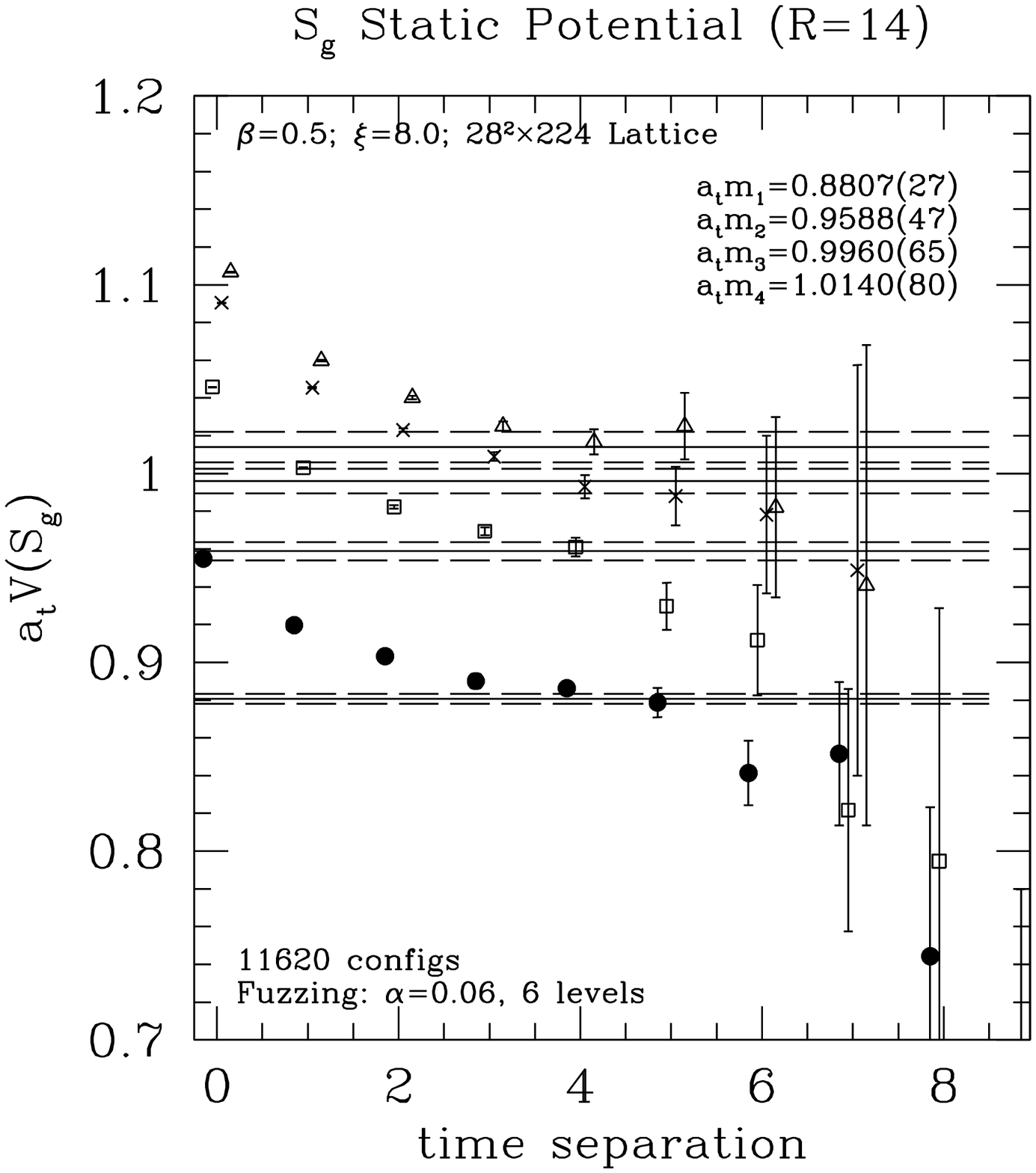}
\caption[cap]{Effective masses and fits for the lowest four energies
of Example 2 using the Bayesian analysis described in 
Sec.~\protect\ref{sec:spectrum} and $0\leq t\leq 9$. 
Prior parameter values are $\Ne=54$, $\varepsilon=0.1$, 
$\sigma_\varepsilon=0.05$, $\Gamma=0.9$, $\sigma_\Gamma=0.2$,
$\gamma=0.05$, $\sigma_\gamma=0.05$.
\label{fig:example2C}}
\end{figure}

\begin{figure}[t]
\renewcommand{\tabcolsep}{1.5mm}
\includegraphics[width=3.0in,bb= 20 90 580 483 ]{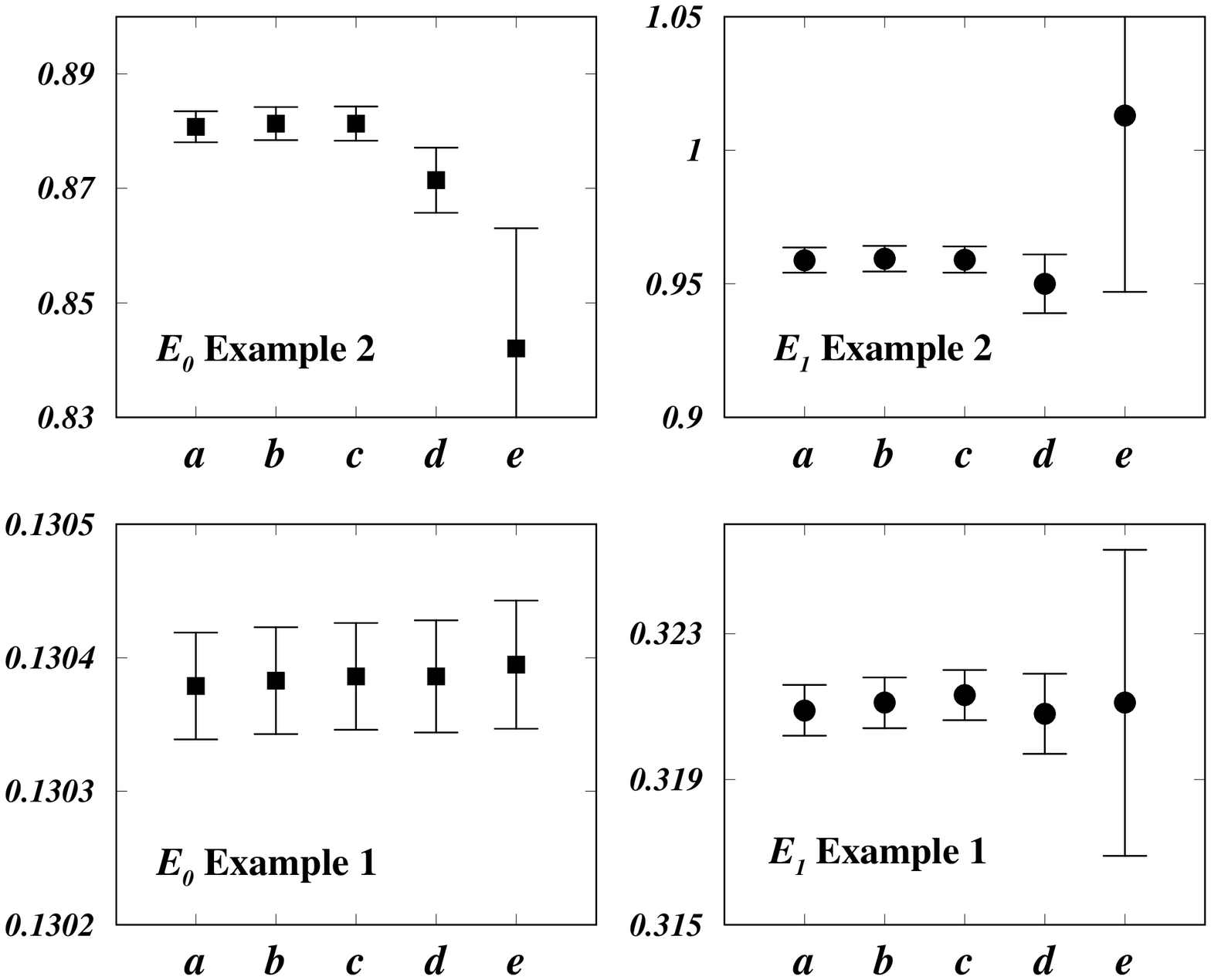}
\caption[cap]{Sensitivity of $E_0$ and $E_1$ to changes in the parameters
of the Bayesian prior.  For each point, the parameters in the prior
are the same as stated in Figs.~\protect\ref{fig:example1B} and
\protect\ref{fig:example2C}, except for the following changes: \\
\begin{tabular}{cllll}
(a)& $\varepsilon=0.1$, & $\sigma_\varepsilon=0.05$, & & \\
(b)& $\varepsilon=0.2$, & $\sigma_\varepsilon=0.10$,  & & \\ 
(c)& $\varepsilon=0.3$, & $\sigma_\varepsilon=0.15$, & & \\ 
(d)& $\Gamma=0.8$, & $\sigma_\Gamma=0.3$, & $\gamma=0.1$, & $\sigma_\gamma=0.1$,\\
(e)& $\Gamma=0.7$, & $\sigma_\Gamma=0.7$, & $\gamma=0.7$, & $\sigma_\gamma=0.7$.
\end{tabular}
\label{fig:priorsens}}
\end{figure}

The results of applying this Bayesian curve fitting method in the
two examples previously described are shown in Figs.~\ref{fig:example1B}
and \ref{fig:example2C}.  The fit values shown correspond to the
mode of each quantity, that is, the value of the quantity at the
maximum of the posterior distribution. The errors have been computed
assuming a normal distribution about this maximum for reasons discussed
below.  These errors automatically combine the statistical errors in the
data with the systematic uncertainties in our prior knowledge.
The method works remarkably well in both examples,
eliminating the need to subjectively determine $t_{\rm min}$.  The
insensitivity of the results to the parameters in the prior probability
is demonstrated in Fig.~\ref{fig:priorsens}.  Notice that the error
estimates increase significantly only in case (e) which is a {\em drastic}
change in the prior.  This insensitivity indicates that $E_0$ and $E_1$ are
largely being determined by the data.  Of course, the data contain
little information about many of the other parameters, and hence, such
parameters depend strongly on the prior.  We have found that this new method
works well as long as $\varepsilon$ is not set much larger than the expected
level spacings in the system and the allowed ranges of the coefficients are
not set too large, as in case (e).  

Currently, there remains one fly in the ointment, namely, the computation of
the error estimates.  We have evaluated the multi-dimensional integrals
needed to calculate the variances in the fit parameters using both the
Metropolis method and HMC and have encountered severely long auto-correlations.
The presence of many poorly constrained parameters leads to long, narrow
ridges in the probability distribution which hamper the
Markov chain Monte Carlo integration.  For the Metropolis method,
we have commonly observed auto-correlation lengths in the tens of thousands.
The problem is somewhat ameliorated with HMC, reducing the auto-correlation
lengths by factors of twenty or so, but clearly, this is still not sufficient.
Work continues on this topic.  An alternative method of computing the errors
(which deviates from a strict Bayesian philosophy) is suggested in
Ref.~\cite{peter}.

\section{CONCLUSION}

Bayesian regression techniques are a viable alternative method for
extracting physical observables from stochastically-determined correlation
functions.  By allowing the regression to take into account prior knowledge
of the system from theoretical considerations and/or previous experience,
Bayesian methods can make use of information obtained at short
temporal separations and can incorporate systematic effects into
error estimates of the observables.  However, Bayesian methods are not
a cure for bad data.

Much of this work arose from a collaborative effort to explore Bayesian
methods. Others involved in this effort are Peter Lepage, Bryan Clark,
Christine Davies, Kent Hornbostel,  Paul Mackenzie,
and Howard Trottier.  The author wishes to acknowledge especially fruitful
discussions with Peter Lepage, and also with Robert Swendsen, Julius Kuti,
Jimmy Juge, David Richards, and Mike Peardon.  This work was supported by
the U.S.\ National Science Foundation under Award PHY-0099450.

\end{document}